\documentclass[aps,prl,twocolumn,showpacs,superscriptaddress, floatfix]{revtex4-1}

\usepackage[utf8x]{inputenc}
\usepackage{hyperref}
\hypersetup{
  colorlinks   = true, 
  urlcolor     = blue, 
  linkcolor    = blue, 
  citecolor   = red 
}
\usepackage{graphicx}

\usepackage{amsmath}
\usepackage{amsfonts}
\usepackage{amssymb}
\newcommand{\unit}[1]{\,\mbox{#1}}
\newcommand{\ket}[1]{\left| #1\right\rangle}

\begin{document}

\title{Storage enhanced nonlinearities in a cold atomic Rydberg ensemble}
\author{E. Distante}
\email{emanuele.distante@icfo.es}
\affiliation{ICFO-Institut de Ciencies Fotoniques, The Barcelona Institute of Science and Technology, 08860 Castelldefels (Barcelona), Spain}
\author{A. Padrón-Brito}
\affiliation{ICFO-Institut de Ciencies Fotoniques, The Barcelona Institute of Science and Technology, 08860 Castelldefels (Barcelona), Spain}
\author{M. Cristiani}
\affiliation{ICFO-Institut de Ciencies Fotoniques, The Barcelona Institute of Science and Technology, 08860 Castelldefels (Barcelona), Spain}
\author{D. Paredes-Barato}
\affiliation{ICFO-Institut de Ciencies Fotoniques, The Barcelona Institute of Science and Technology, 08860 Castelldefels (Barcelona), Spain}
\author{H. de Riedmatten}
\affiliation{ICFO-Institut de Ciencies Fotoniques, The Barcelona Institute of Science and Technology, 08860 Castelldefels (Barcelona), Spain}
\affiliation{ICREA-Institució Catalana de Recerca i Estudis Avançats, 08015 Barcelona, Spain}

\date{\today}
 \begin{abstract}
The combination of electromagnetically induced transparency (EIT) with the nonlinear interaction between Rydberg atoms provides an effective interaction between photons. In this paper, we investigate the storage of optical pulses as collective Rydberg atomic excitations in a cold atomic ensemble. By measuring the dynamics of the stored Rydberg polaritons, we experimentally demonstrate that storing a probe pulse as Rydberg polaritons strongly enhances the Rydberg mediated interaction compared to the slow propagation case. We show that the process is characterized by two time scales. At short storage times, we observe a strong enhancement of the interaction due to the reduction of the Rydberg polariton group velocity down to zero. For longer storage times, we observe a further, weaker enhancement dominated by Rydberg induced dephasing of the multiparticle components of the state. In this regime, we observe a non-linear dependence of the Rydberg polariton coherence time with the input photon number. Our results have direct consequences in Rydberg quantum optics and may enable the test of new theories of strongly interacting Rydberg systems. 
\end{abstract}

\maketitle

The possibility to control the interaction between photons provided by highly nonlinear media is a key ingredient to the goal of quantum information processing (QIP) using photons and a unique tool to study the dynamics of many-body correlated systems \cite{Chang14}. Many different systems showing high nonlinear optical response at the single-photon level have been studied during the past years ranging from resonators coupled to single atoms \cite{Dayan2008,Reiserer2013,Reiserer2014,Tiecke2014,Shomroni2014}, atomic ensembles \cite{Chen2013}, to artificial two-level atoms \cite{Fushman2008,Volz2012}.
 
A promising strategy to perform different QIP tasks using photons as carriers is the combination of electromagnetically induced transparency (EIT) \cite{HAU1999,Fleischhauer00,Fleischhauer02,Fleischhauer05} and Rydberg atoms \cite{Gallagher05} (see for example \cite{Friedler05,Shahmoon11,Gorshkov10,Gorshkov11, He11,Gorshkov12,Paredes14,He14,Khazali15,Das15,Pritchard10,Dudin12,Peyronel12,Firstenberg2013,Maxwell13,Tiarks14,Baur14,Ding2015,Gorniaczyk14,Gorniaczyk2016,Boddeda15,Tiarks16}). Using EIT one maps the state of the photons into atomic coherence in the form of Rydberg dark-state polaritons (DSPs) by means of an auxiliary coupling field. The strong Rydberg dipole-dipole (DD) interaction between neighboring excitations shifts the multiply-excited states from being resonantly coupled when these excitations are closer than a certain length called the \emph{blockade radius}, $r_{\rm b}$ \cite{Pritchard10}. This way, only a single excitation can be created inside of a blockaded volume of the atomic cloud (so-called \emph{super-atom}). This phenomenon, known as \emph{Rydberg blockade}, has been used in combination with EIT to generate quantum states of light \cite{Peyronel12,Firstenberg2013,Maxwell13}, single-photon switches and transistors \cite{Tiarks14,Baur14,Gorniaczyk14,Gorniaczyk2016} as well as a $\pi$ phase shift controlled with single-photon level pulse \cite{Tiarks16}. These experiments typically require very high atomic densities and high-lying Rydberg states. By switching off and back on the coupling field, photons can be stored as Rydberg excitations and retrieved at later time \cite{Dudin12,Maxwell13}. In this case the DD interaction dephases the collective emission of the multiparticle components of the stored photonic states \cite{Bariani12b, Bariani12}. This feature was used to implement a deterministic single-photon source \cite{Dudin12}. 

The key point of all these experiments is the strong nonlinear response arising from the DD interaction between high-lying Rydberg states. In the present paper, we experimentally demonstrate that storing the input photons as Rydberg excitations strongly enhances the nonlinear interaction when compared to the propagation case. The profound difference between propagating and storing Rydberg DSPs and its application in many-body Rydberg physics and QIP has been recently theoretically discussed \cite{Bariani12b,Moos15}. We show experimentally that the underlying many-body dynamics of strongly interacting DSPs during storage is characterized by two different time scales. A strong enhancement of the interaction happens at time scales shorter than what can be measured in this experiment. At longer time scales, the dynamics is dominated by the dephasing of multiparticle components of the input states. We confirm the latter by measuring for the first time the nonlinear dependence of the coherence time of stored Rydberg DSPs with respect to the input photon number \cite{Bariani12}. 
Our results have a direct consequence in Rydberg quantum optics, demonstrating that the regime of strongly interacting DSPs required in most of the protocols can be achieved by storing the the light even for a very short time. Moreover our experiment is a step forward in understanding the complicated many-body physics of the strongly interacting DSPs during storage \cite{Moos15}.

 \begin{figure*}[t]
 \includegraphics[width=\linewidth]{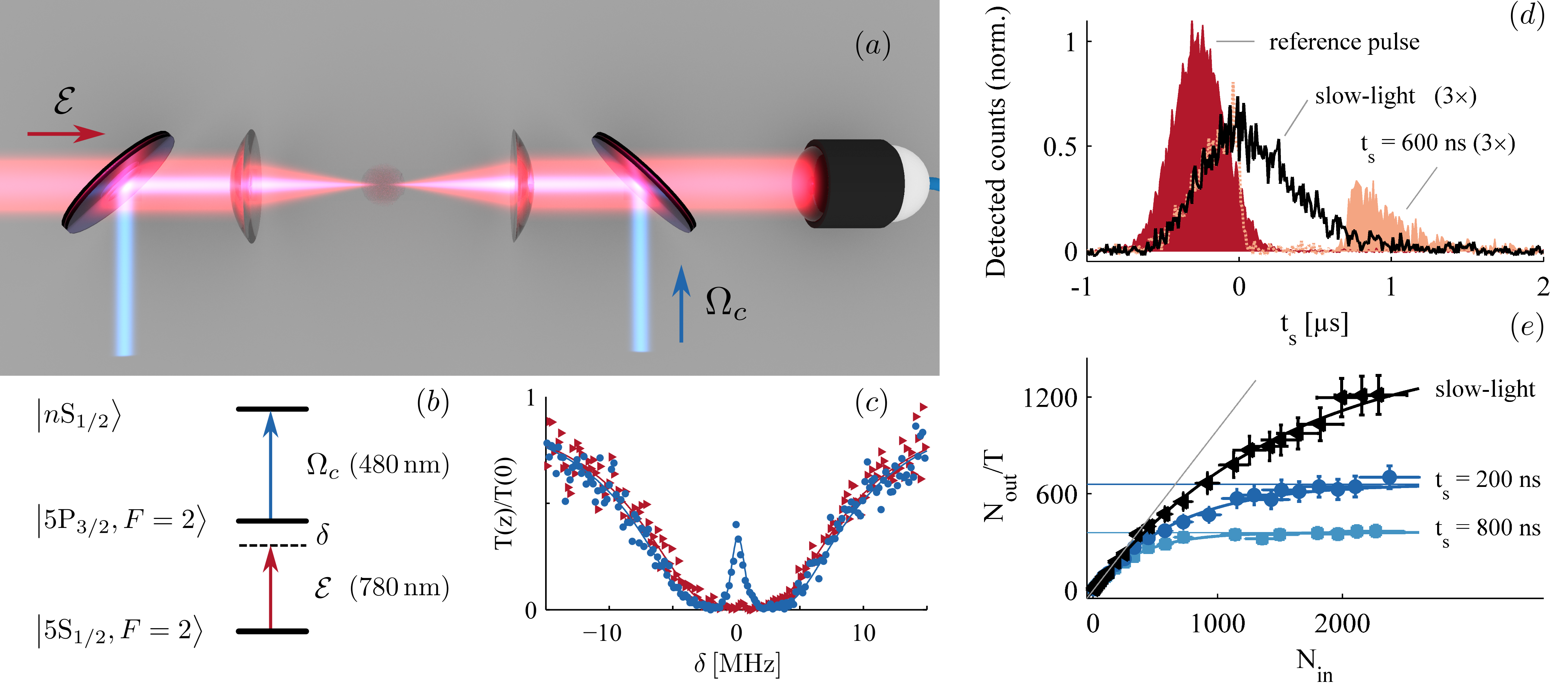}
 \caption{(a) Counterpropagating probe (red) and coupling (blue) beams are focused using aspheric lenses onto a cold cloud of $^{87} \rm Rb$ atoms. Probe and coupling beams are combined and separated using dichroic mirrors. Probe photons are detected using a fiber coupled single-photon APD. (b) Simplified atomic level scheme. (c) Probe transmission traces when coupling beam is off (red triangles) and on, showing the typical EIT transparency window (blue circles). Solid lines are fits to the data \cite{Xiao1995} (d) Normalized and background subtracted counts of an input Gaussian probe (red area) when propagating as slow-light DSP (black line) and when stored for $t_{\rm s} = 600\,{\rm ns}$ (orange area). Here the input photon number is $N_{\rm in} = 23.2 \pm 1.2$, with efficiencies $\eta = 0.336 \pm 0.006$ and $\eta = 0.078 \pm 0.002$ for the slow- and stored light respectively. Dashed orange line represents the leaked pulse during the storage process. (e) $N_{\rm out}$ normalized by the linear process efficiency $T$ as a function of the input photon number $N_{\rm in}$ for the slow-light case (black triangles) and for two storage times. Solid curves represent fits with Eq.\eqref{model}. Straight lines represent the linear behavior $N_{\rm out}/T = N_{\rm in}$ (oblique) and the saturation level $N_{\rm max}$ (horizontal). The Rydberg state used in these plots is $\ket{70S_{1/2}}$.} 
 \label{fig:fig1}
 \end{figure*} 
Our measurement can be summarized as follows: we send coherent probe pulses with varying intensity and measure the number of emerging photons in the slow-light and in the storage case. The Rydberg DD interaction causes a non-linear input-output intensity relation, eventually leading to the saturation of the output photon number. Stronger, non-linear interactions lead to a reduction of the maximum number of photons sustained by the medium \cite{Baur14}. In a first experiment, we measure $N_{\rm max}$, the saturation plateau normalized by the linear process efficiency, $T$. We show that storage leads to a strong suppression of $N_{\rm max}$ compared to the slow-light case, thus demonstrating a strong enhancement of the Rydberg mediated photon interaction. The dependence of $N_{\rm max}$ on the storage time $t_{\rm s}$ shows that strong suppression happens at a short time scale. In a second experiment, we measure the coherence time of the storage process as a function of the input photon number. We show that higher intensity input fields suffer from stronger dephasing due to the Rydberg DD interaction. 

\paragraph{Experiment} In  Fig.\ref{fig:fig1} a schematic of this system is shown. We probe a cold cloud of $^{87}{\rm Rb}$ atoms using 780 nm light ($\mathcal{E}$) with a detuning $\delta$ with respect to the $\ket{g}\leftrightarrow\ket{e}$ transition, where $\ket{g}=\ket{5{\rm}S_{1/2},F=2}$ and $\ket{e}=\ket{5{\rm}P_{3/2},F=2}$. The atomic sample is obtained using a magneto-optical trap, which generates a cloud of size $\sigma\sim0.8\,\textrm{mm}$ with a peak density $\rho_0 = 3.2 \cdot 10^{10} \unit{cm}^{-3}$, and a temperature $T \sim 87.5\, \mu {\rm K}$ (measured by fluorescence imaging). A strong coupling field at 480 nm light is sent counterpropagating with respect to the probe. Using an excited-state locking scheme \cite{Abel09}, we lock the coupling beam resonantly to the $\ket{e}\leftrightarrow\ket{r} = \ket{n{\rm} S_{1/2}}$ transition, where $n$ is the principal quantum number. The probe and coupling laser fields are focused to waist radii $(w_p, w_c) \approx (7\,\mu{\rm m}, 13\,\mu{\rm m})$. This geometry gives $\approx3.9\times 10^{4}$ atoms in the interacting region. The optical depth (OD) of the cloud and the coupling Rabi frequency $\Omega_c$ are extracted by fitting the transmission of the probe as a function of the probe detuning $\delta$ with respect to the  $\ket{5{\rm S}_{1/2}, F=2} \leftrightarrow \ket{5{\rm P}_{3/2} , F=2}$ transition using the model presented in \cite{Xiao1995}. We set them to be OD $\sim 6.2$ and $\Omega_c = (4.38 \pm 0.04)\, \textrm{MHz}$  (see Fig.\ref{fig:fig1}(c)). The probe and the coupling beam are opposite circularly polarized \cite{Low2012a} and the magnetic field is set to zero at the position of the cloud \cite{BorisPhD}. 

When $\delta = 0 $ the presence of the coupling field converts the probe photons into propagating Rydberg DSPs \cite{Fleischhauer00,Fleischhauer02,Fleischhauer05}. These DSPs travel at reduced group velocity $v_{\rm g} \sim |\Omega_c|^2/ (g^2 \rho_0)$, where $g$ is the coupling strength between the probe and the $\ket{g}\leftrightarrow\ket{e}$ transition \cite{Fleischhauer00,Liu2001}. By adiabatically switching off the coupling beam we store the state of the input field as an atomic coherence. The stored excitation is retrieved after a storage time $t_{\rm s}$ by switching the coupling beam back on.
 \begin{figure}
 \includegraphics[width=\columnwidth]{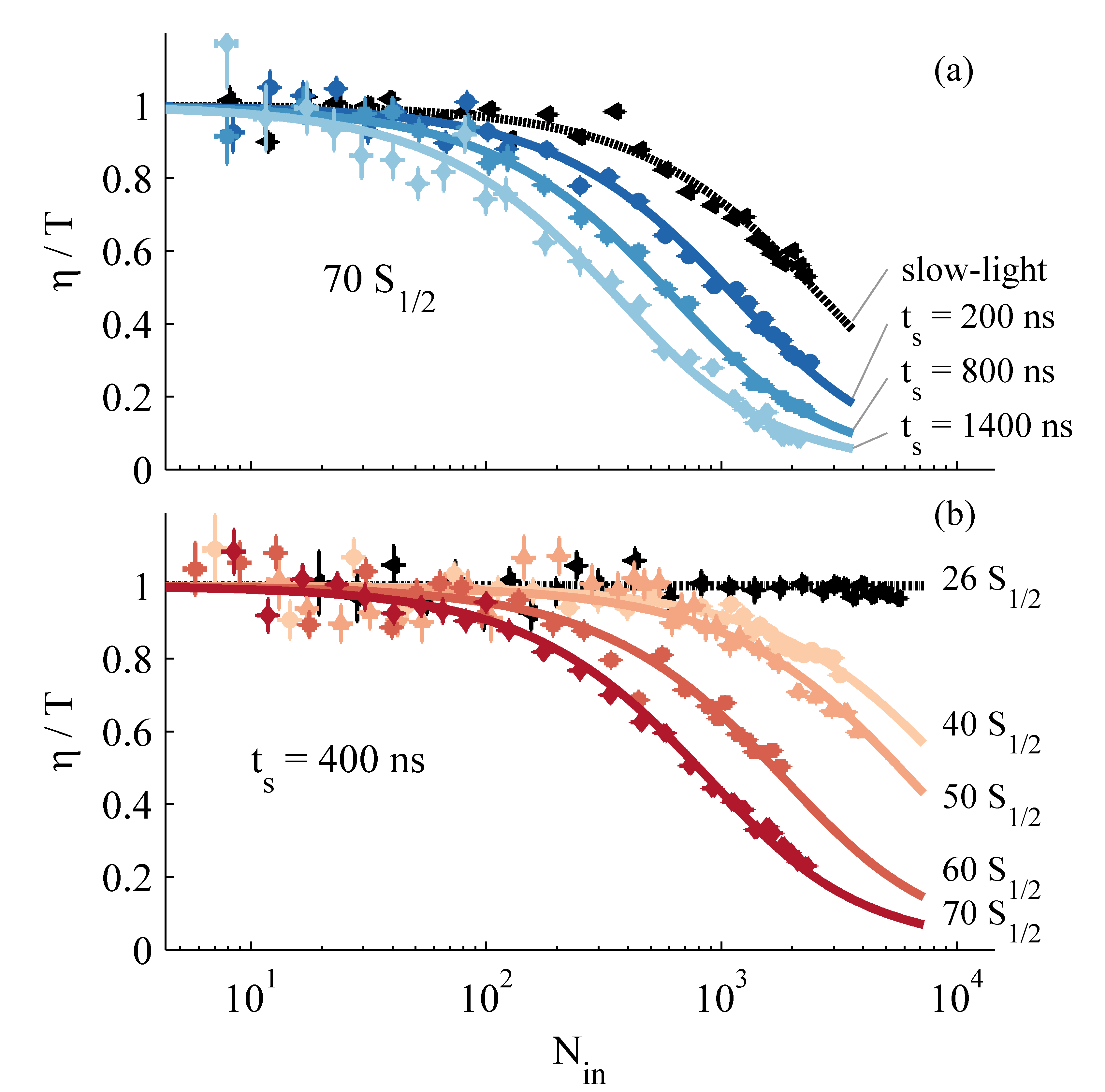}
 \caption{Normalized efficiency $\eta/T$ as a function of the input photon number $N_{\rm in}$ (a) For fixed Rydberg state $\ket {70S_{1/2}}$, comparison between slow-light case (black triangles) with the storage case for different storage times. (b) For fixed storage time $t_{\rm s} = 400 \, \rm ns$ comparison between non-interacting low-lying Rydberg state $26S_{\rm 1/2}$ (black triangles) with stronger interacting, higher $n$ Rydberg states. In both plots, lines are fits with the Eq.\eqref{model}.}  
 \label{fig:fig2}
 \end{figure}  
 \begin{figure}[t]
 \includegraphics[width=\columnwidth]{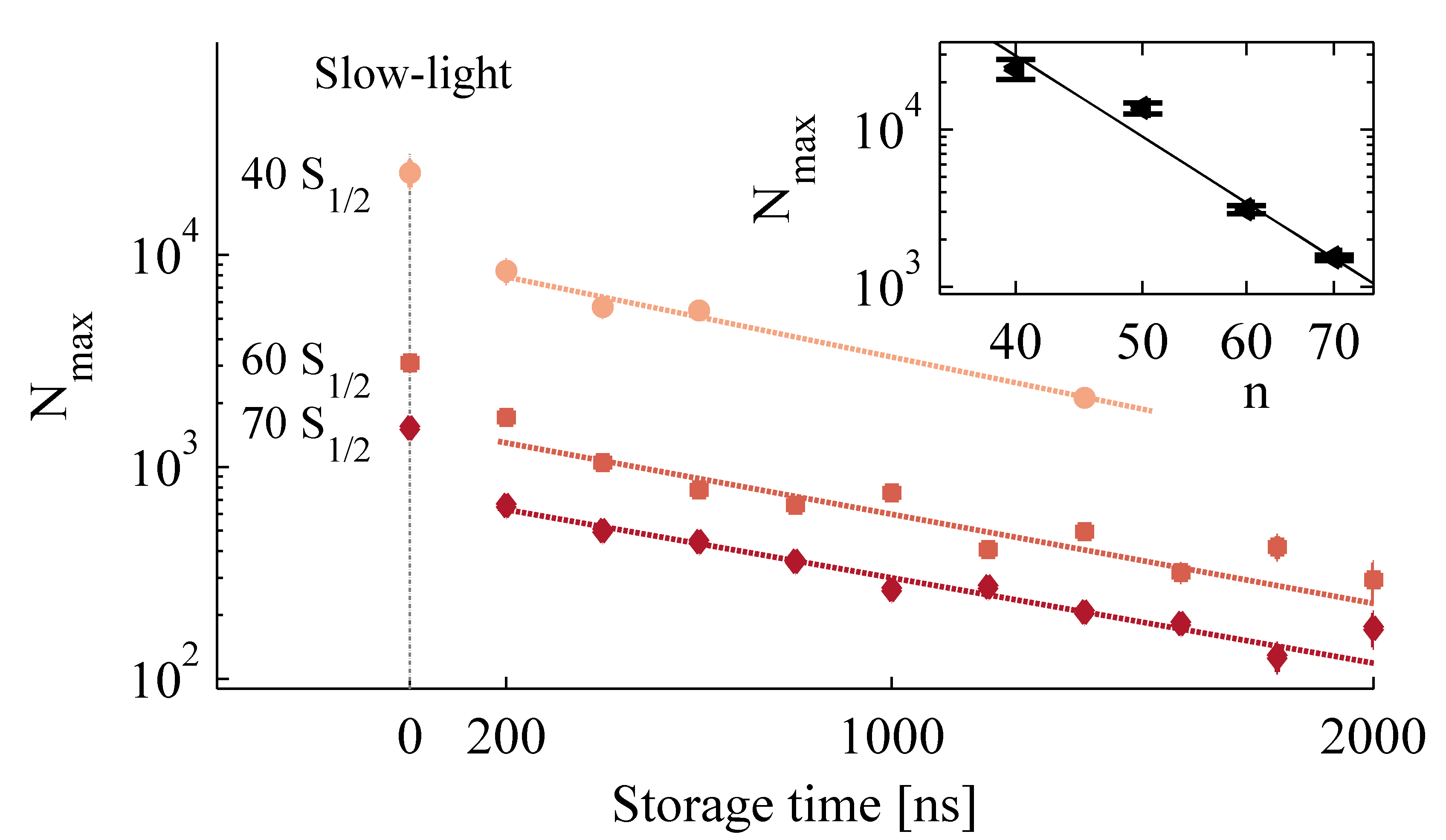}
 \caption{Maximum number of retrieved photons normalized by the process efficiency $N_{\rm max}$ (extracted from the fit with Eq.\eqref{model}) as a function of the storage time $t_{\rm s}$ for different Rydberg states. Comparison with the slow-light case (points at $t_{\rm s} = 0 \,\rm ns$) shows a strong reduction of $N_{\rm max}$ when storage is performed. Dotted lines are exponential fits to the storage data. (Inset) $N_{\rm max}$ as a function of the principal quantum number $n$ for slow-light. Solid line is fit with the function $N_{\rm max} = \alpha n^{-\gamma}$ giving $\gamma = 5.3 \pm 0.2$ (see main text). }  
 \label{fig:fig3}
 \end{figure}  
 
We send a Gaussian, coherent probe pulse with a duration of  $410\unit{ns}$ $({\rm FWHM_{in}})$ and average number of photons $N_{\rm in}$ through the cold atomic gas. 
The light is detected after the ensemble with a single-photon APD and the counts are background subtracted and corrected for detection efficiency. Initially we calibrate $N_{\rm in}$ by measuring the transmitted pulse without loading the atoms. Then we measure $N_{\rm out}$ either when the probe pulse propagates as slow Rydberg DSPs or when they are stored in the $\ket{nS_{1/2}}$ state (see Fig.\ref{fig:fig1}(e)). In the absence of Rydberg interaction, the average number of photons $N_{\rm out}$ in the emerging pulse increases linearly with $N_{\rm in}$,  $N_{\rm out} =  T N_{\rm in}$. Here $T<1$ represents imperfect process efficiency. During slow-light propagation, this is caused by the decoherence of the ground Rydberg transition, which includes the natural lifetime of the Rydberg state, atomic collisions, coupling with external fields and laser linewidth. The storage process efficiency is further limited by imperfect pulse compression inside the medium (due to low OD) and by the dephasing of the collective state during the storage time, which is dominated by coupling with external fields and atomic motion. 
When the number of input photons is increased a nonlinear dependence arises, eventually leading to saturation of $N_{\rm out}$.  To quantify the effective interaction we fit our data with the model proposed in \cite{Baur14}. In that model the input-output relation is described by:
\begin{equation}\label{model}
N_{\rm out} = N_{\rm max} T(1-e^{-N_{\rm in}/N_{\rm max}} ) 
\end{equation}
where $T$ represents the linear efficiency of the process  at low photon number and $N_{\rm max}$ is the maximum number of photons that can emerge from the medium when unitary efficiency $T = 1$ is considered. As explained in \cite{Baur14}, $N_{\rm max}$ decreases for stronger Rydberg interaction and can be used to quantify the effective blockade of the output field. Fig.\ref{fig:fig1}(e) reports an example of the data for $n=70$ together with the fit using Eq.\eqref{model}. All the data presented in this manuscript can be found in \cite{DataZenodo}. 

In Fig.\ref{fig:fig2}(a) a re-scaled efficiency $\eta /T$ (being $\eta=N_{\rm out}/N_{\rm in}$) is shown for the high-lying state $\ket {70S_{1/2}}$ at different storage times. The data show that $\eta/T$ tails off at lower $N_{\rm in}$ for longer storage times as a consequence of stronger nonlinearity. Similar data have been taken for a variety of Rydberg states (see e.g. Fig.\ref{fig:fig2}(b) for the results with $t_{\rm s}=400\,{\rm ns}$) and the fit results are shown in Fig.\ref{fig:fig3}. One could argue that saturation may arise as a result of medium saturation, when the density of photons and atoms inside the medium are comparable. However, in Fig.\ref{fig:fig2}(b) we observe that the response of the medium is linear (that is, the efficiency does not depend on $N_{\rm in}$) at low-lying Rydberg states, where the interaction is negligible \cite{Supp}\nocite{Han16,Gorshkov07,Chen13,Hsiao16,Firstenberg2016}. 

In Fig.\ref{fig:fig3} we show $N_{\rm max}$ for different Rydberg levels, both in the slow-light or in the storage case. As expected, when $n$ is increased $N_{\rm max}$ is reduced, due to stronger Rydberg interaction. In the propagation case this can be understood as a consequence of the blockade effect. When the density of photons in the medium becomes comparable to the density of \textit{super-atoms} $\rho_{\rm SA}=3/4\pi r_{\rm b}^3$, the medium saturates \cite{Petrosyan11}. Since $r_{\rm b} \sim n^{11/6}$, this condition is reached at a lower number of photons for a higher Rydberg state. Following this, the maximum number of photons supported by the medium scales as $ N_{\rm max} \sim n^{-11/2}$. The inset in Fig.\ref{fig:fig3} shows a fit for the slow-light case with the function $N_{\rm max} = \alpha n^{-\gamma}$ which gives $\gamma = 5.3 \pm 0.2$. 

When the probe pulse is stored, saturation occurs at a lower number of photons (an order of magnitude difference for $t_{\rm st}=2\,\mu s$), as shown in Fig.\ref{fig:fig3}. Data show that  $N_{\rm max}$ is strongly reduced soon after storage is performed. The two time scales of the process are evident when noticing that even an exponential fit of $N_{\rm max}$ in the storage case (dotted lines in Fig.\ref{fig:fig3}) fails to include the data of the slow-light case, represented by the $t_{\rm s}=0 \, \rm ns$ points in Fig.\ref{fig:fig3}. In the ideal limit of zero decoherence between the ground and the Rydberg state, the blockade radius increases without bounds when $\Omega_c$ goes to zero, according to the naive formula for the blockade radius $r_{\rm b} = \sqrt[6]{C_6/\delta_{\rm EIT}}$ in terms of the EIT bandwidth $\delta_{\rm EIT} \propto \Omega_c^2$ and the van der Waals coefficient $C_6$ \cite{Moos15}. This is not consistent with our data nor with other experimental results \cite{Maxwell13, Maxwell14}. A recent description by Moos \textit{et al.} \cite{Moos15} suggests that $\Omega_c$ has to be replaced with $\Omega_{eff}^2 = g^2 \rho_0+\Omega_c^2$ upon storage. According to this new description, the blockade radius during storage becomes $r_{\rm b} = \sqrt[6]{C_6\Gamma/g^2 \rho_0}$. As a consequence, the blockaded volume would not increase significantly during the storage process and it could not be used to understand the data. Nevertheless Moos \textit{et al.} suggest that the strongly interacting regime is achievable when the ratio between the Rydberg interaction and the kinetic energy of the DSP is strongly increased; this regime is achieved during the storage process when $v_{\rm g}$ is reduced to zero. This theory also suggests other specific effects (such as a quasicrystalline density of stored photons) which are interesting but not within the reach of our current setup.
 \begin{figure}[t]
 \includegraphics[width=\columnwidth]{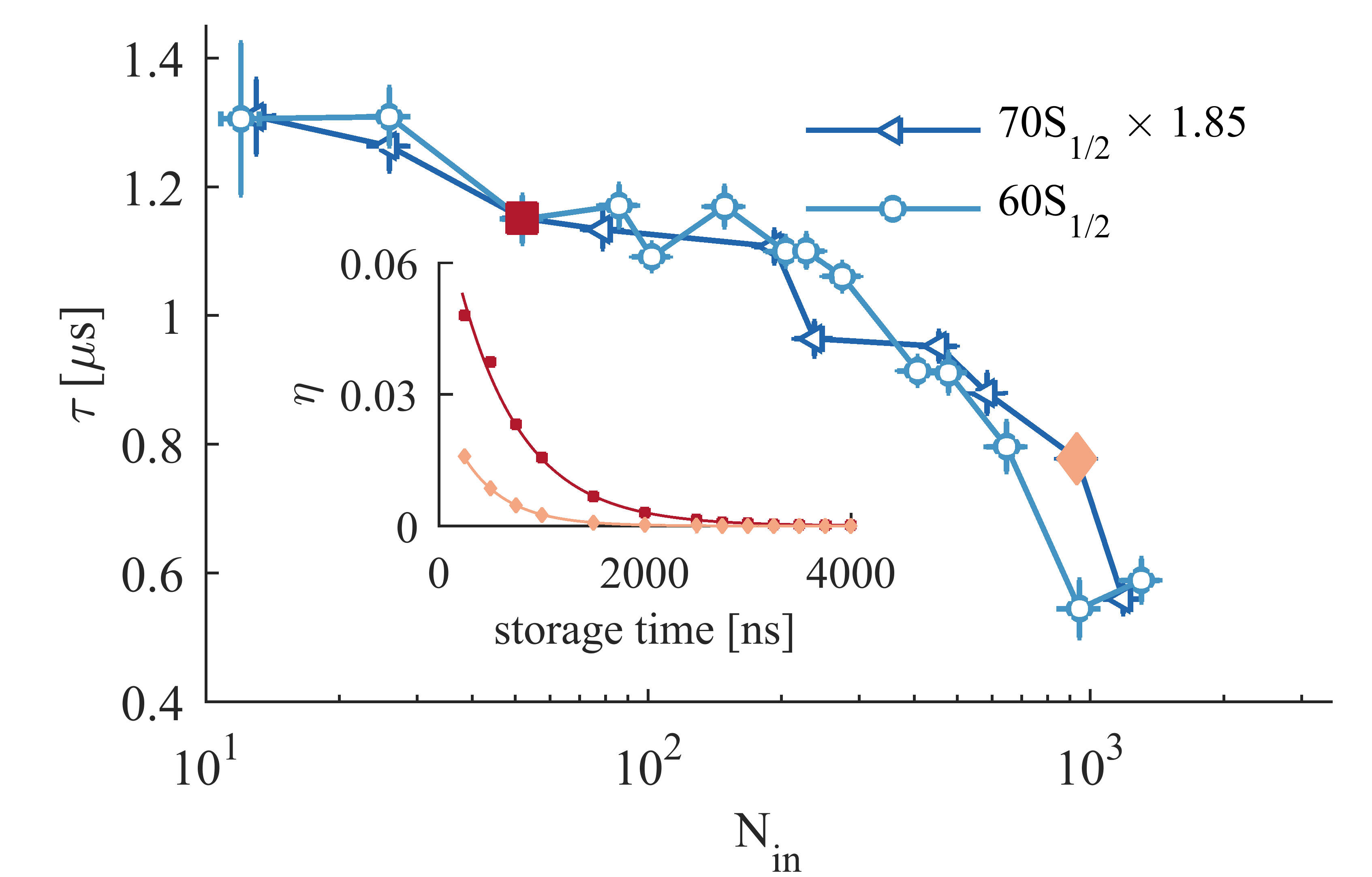}
 \caption{(Inset) Storage efficiency $\eta$ as a function of storage time $t_{\rm s}$ for input photon number $N_{\rm in} \sim 52$ (squares) and $N_{\rm in} \sim 934$ (diamonds). Solid lines represent fits using an exponential function. (Main Plot) Coherence time $\tau$ extracted from the fit as a function of the input photon number $N_{\rm in}$ for Rydberg levels $n=60,\,70$ (empty triangles and circles respectively). Filled squares and diamonds in the $n = 70$ set represent the fit of the data shown in the inset.}  
 \label{fig:fig4}
 \end{figure}  

At longer time scales, the Rydberg DD interaction acts as an extra source of dephasing for the many-body components of the stored DSP effectively blocking the collective emission of such components in the retrieved mode. This effect \cite{Bariani12b} has been observed before and it has been exploited to generate single photons deterministically \cite{Bariani12,Dudin12}. Here we show the first detailed time-dependent study. We measure the storage efficiency $\eta$ versus the storage time $t_{\rm s}$ for a variety of input photon numbers $N_{\rm {in}}$. The inset of Fig.\ref{fig:fig4} reports an example of the efficiency data for the $\ket{70S_{1/2}}$ state for two different $N_{\rm in}$. We extracted the 1/e coherence time $\tau$  by fitting $\eta(t_{\rm s})$ with a model shown in \cite{Supp}. The results are summarized in Fig.\ref{fig:fig4} where we show how $\tau$ depends on $N_{\rm in}$ for two different Rydberg states $\ket {60S_{1/2}}$ and $\ket {70S_{1/2}}$. At low photon numbers, we observe larger dephasing at higher principal quantum number, likely due to stray external electric field \cite{Supp}. At higher $N_{\rm in}$, the interaction between Rydberg states introduces another source of dephasing, resulting in a reduction of $\tau$. Both Rydberg states show similar dependence of $\tau$ (when normalized at a low number of photons, as shown in Fig.\ref{fig:fig4}) with respect to $N_{\rm in}$. At first surprising, this result can be understood by noticing that the system starts to evolve from a partially blockaded configuration, contrary to the situation studied in \cite{Bariani12b,Dudin12}. Following theory presented in \cite{Bariani12b}, the interaction between Rydberg states induces a phase shift on the \textit{m}-body component of the storage state $\phi_{\mu_1\dots\mu_m}  =-t\sum_{1\leqslant i < j \leqslant m} V_{\mu_i\mu_j}/\hbar$. Here $V_{\mu_i\mu_j}$ is the the Van der Waals potential describing the interaction between two Rydberg excitations $\mu_{i}$ and $\mu_{j}$, which is strongly state dependent. Nevertheless, due to the blockade effect, two excitations cannot be closer than $r_{\rm b}$. At this distance, the dipole potential is fixed by the EIT linewidth: $V(r_{\rm b}) = \hbar \delta_{\rm EIT}$. Since $\delta_{\rm EIT}$ is similar in the two Rydberg states considered in our experiment, we expect both states to present similar dephasing rates. For the non-interacting case ($n = 26$) we do not observe any changes of $\tau$ as a function of $N_{\rm in}$ \cite{Supp}.

\paragraph{Conclusions} We have performed the first extensive measurement of the dynamics of stored Rydberg DSPs. Our data clearly demonstrate that storing photons as a Rydberg DSP enhances the Rydberg mediated interaction when compared to the slowly propagating case. This result may open the door to obtaining strong photon-photon interactions at moderate atomic densities and lower Rydberg states. Our results, combined with efficient storage \cite{Chen13,Hsiao16}, would facilitate photonic QIP using Rydberg atoms by relaxing the stringent requirements of high densities \cite{Balewski2014,Firstenberg2016} and high Rydberg levels to enhance the interactions between polaritons. We have discussed the many-body dynamics of the process, showing that two different time scales are present. We suggest that a recent theory proposed by Moos \textit{et al.} in \cite{Moos15} might explain our results at short time scales. At long time scales, we have presented the first time-dependent measurement of the dephasing induced by the Rydberg DD interaction and we have shown its clear dependence on the input photon number. In the future, our data might allow to test more detailed models of interacting Rydberg DSPs, shedding light on the strongly interacting many-body physics with Rydberg atoms. On the experimental side, reducing the lasers linewidth and the cloud temperature would enable the study of the dephasing at longer storage times and at higher Rydberg levels. Finally, these results can be extended to show nonlinearities at the single-photon level by increasing the density of the cloud and by reducing the size of the sample.

\begin{acknowledgments}
	We thank M. Fleischhauer, M. Moos, R. Unanyan, C. Simon and M. Khazali for useful discussions. DPB has received funding from the European Union’s Horizon 2020 research and innovation programme under the Marie Skłodowska-Curie grant agreement No 658258. We acknowledge financial support by  the ERC starting grant QuLIMA, by the Spanish Ministry of Economy and Competitiveness (MINECO) Severo Ochoa through grant SEV-2015-0522, by AGAUR via 2014 SGR 1554 and by Fundació Privada Cellex.
\end{acknowledgments}

\bibliography{enhancedNonlinearities_arxiv}
\clearpage
\onecolumngrid
\section{Supplemental Material for Storage enhanced nonlinearities in a cold atomic Rydberg ensemble}
\twocolumngrid
\section{Rydberg EIT in the blockaded regime}

When performing EIT experiment involving Rydberg states, one observes a reduction of the transparency peak as a function of the input photon rate due to the Rydberg blockade effect. The blockade mechanism can be distinguished from other mechanisms (such as dephasing or ion-formation) by looking at the characteristics of the EIT transparency peak at different photon rates. 

In this section we show the results of our EIT experiment varying the input photon rate when the coupling beam is tuned to the $\ket{5P_{3/2}}\rightarrow\ket{60S_{1/2}}$ transition, similarly to what has been shown in \cite{Pritchard10}. In order to extract the resonance frequency and the width of the EIT peak, we fit the data of the transmitted probe photon as a function of the probe detuning $\delta$ with respect to the transition $\ket{5S_{1/2}}\rightarrow\ket{5P_{3/2}}$ with the function
\begin{equation}\label{eit:eq}
T(\delta) = T_{\rm 2L}(\delta) + T_{0} \, \rm exp\left(-\rm ln(2)\frac{4(\delta-\delta_0)^2}{\rm \Delta_{\rm EIT}^2}\right)
\end{equation} 

where $\delta_{0}$ is the EIT peak resonance frequency, $T_{0}$ is the transmission at $\delta=\delta_{0}$, $\rm \Delta_{\rm EIT}$ is the full width at half maximum of the EIT peak and $T_{\rm 2L}(\delta)$ is the transmission as a function of the probe detuning $\delta$ for a two-level system, defined as:

\begin{equation}\label{twolevel:eq}
T_{\rm 2L}(\delta) = \rm exp\left( -\frac{\rm OD }{1+4(\delta/\Gamma)^2}\right)
\end{equation}

where now OD is the optical depth and $\Gamma$ is the natural linewidth of the $\ket{5S_{1/2}}\rightarrow\ket{5P_{3/2}}$ transition. 

\begin{figure}[h]
	\includegraphics[width=\linewidth]{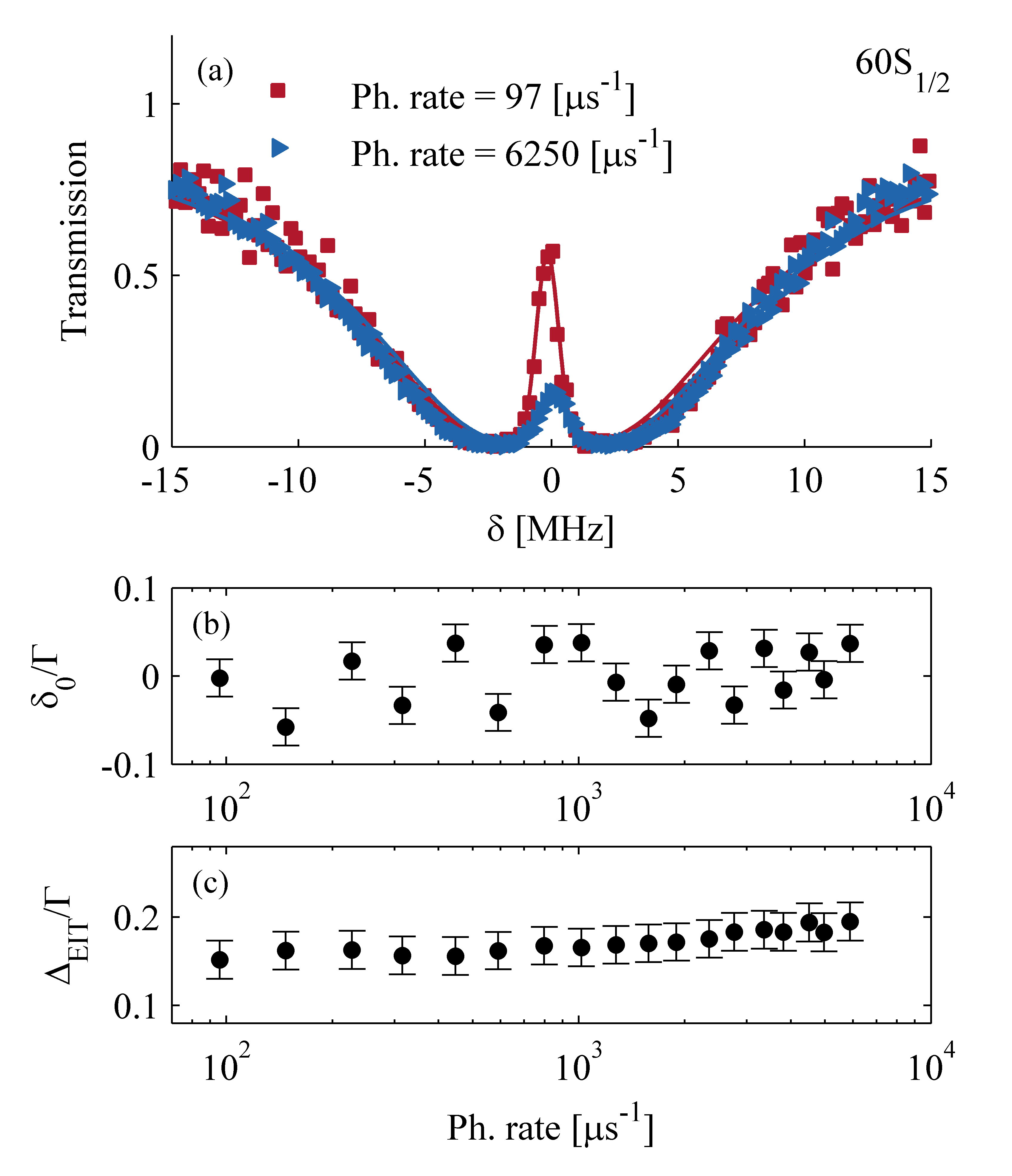}
	\caption{(a) Examples of EIT transmission through the sample as a function of probe laser detuning for two different input photon rates in the level $60S_{1/2}$. Solid lines are fits with the function in Eq. \eqref{eit:eq}. (b) The EIT resonance frequency $\delta_{0}$ and (c) the full-width at half maximum $\Delta_{\rm EIT}$ of the EIT transparency window as a function of the input photon rate.}  
	\label{fig:eit}
\end{figure} 

As shown in Fig. S\ref{fig:eit}, we observe that $\delta_{0}$ and $\Delta_{\rm EIT}$ remain constant with increasing photon numbers. This is consistent with the observation of Rydberg blockade in our (large) sample \cite{Pritchard10, Han16}.

\section{Linear case: $26S_{1/2}$}
\begin{figure}[h]
	\includegraphics[width=\linewidth]{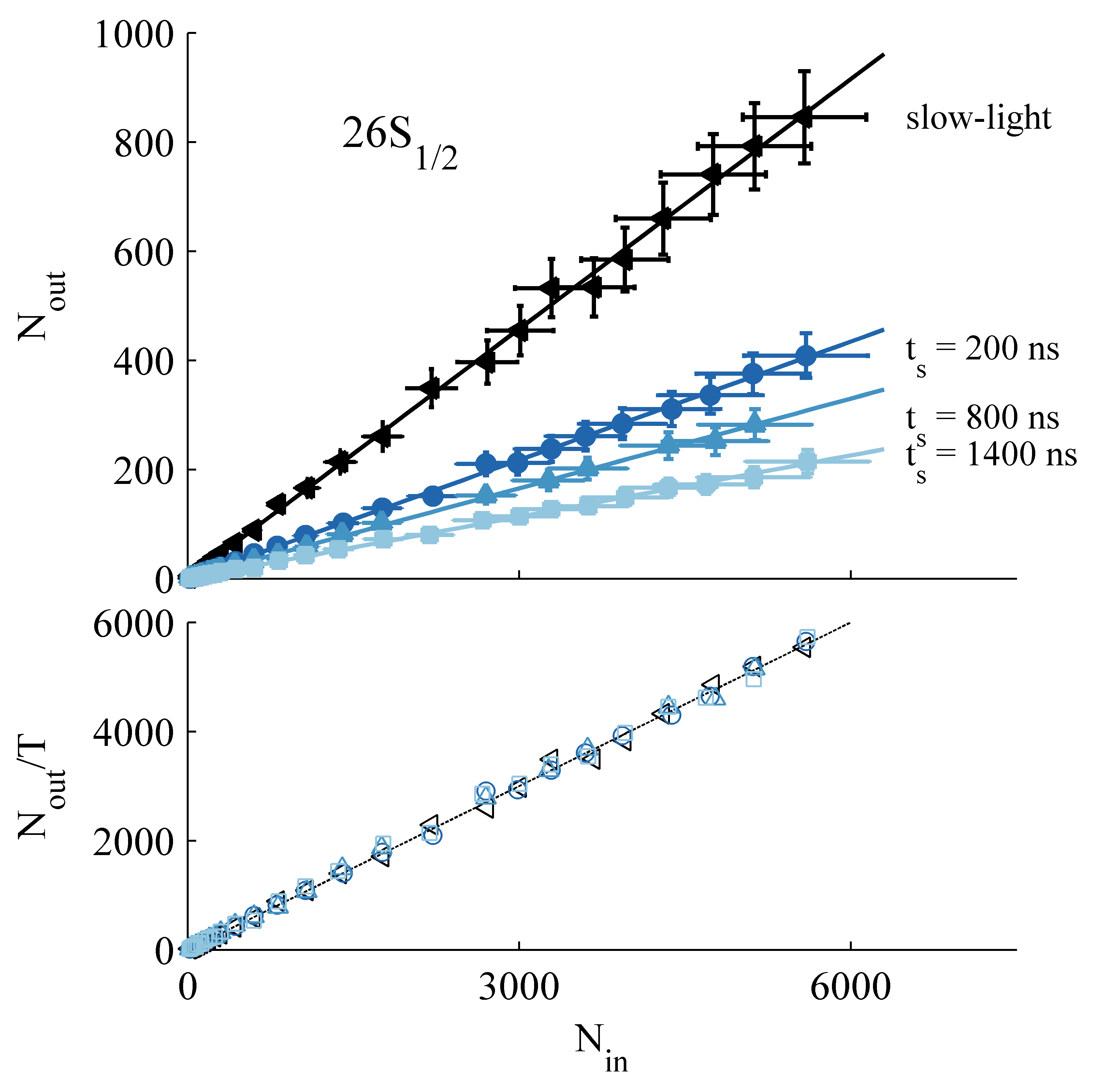}
	\caption{(Top) $N_{\rm out}$ as a function of $N_{\rm in}$ for the $26S_{1/2}$ Rydberg level for slow-light (black) and different storage times (shades of blue), showing a linear behavior (lines). (Bottom) When the number of output photons is divided by the linear efficiency T, all of the points collapse into a single line, showing that the process is linear and independent on the storage time.}  
	\label{fig:26S}
\end{figure}

For low-lying Rydberg states we expect dipole-dipole interactions to be irrelevant at typical interatomic distances for our density. This implies that both slow-light propagation and storage are linear processes, and no saturation should be present. In Fig. S\ref{fig:26S} we show $N_{\rm out}$ as a function of $N_{\rm in}$ for the Rydberg state $26S_{1/2}$ both for the slow-light case and for different storage time. Our data show clearly a linear input-output relation for this level and no saturation either for the slow-light case and for all the storage times considered. 

At higher input photon numbers, it might be possible to observe saturation when the number of photons in the medium is comparable to the number of atoms which is of the order of few $10^4$ in our interaction region.

\begin{figure}[h]
	\includegraphics[width=\linewidth]{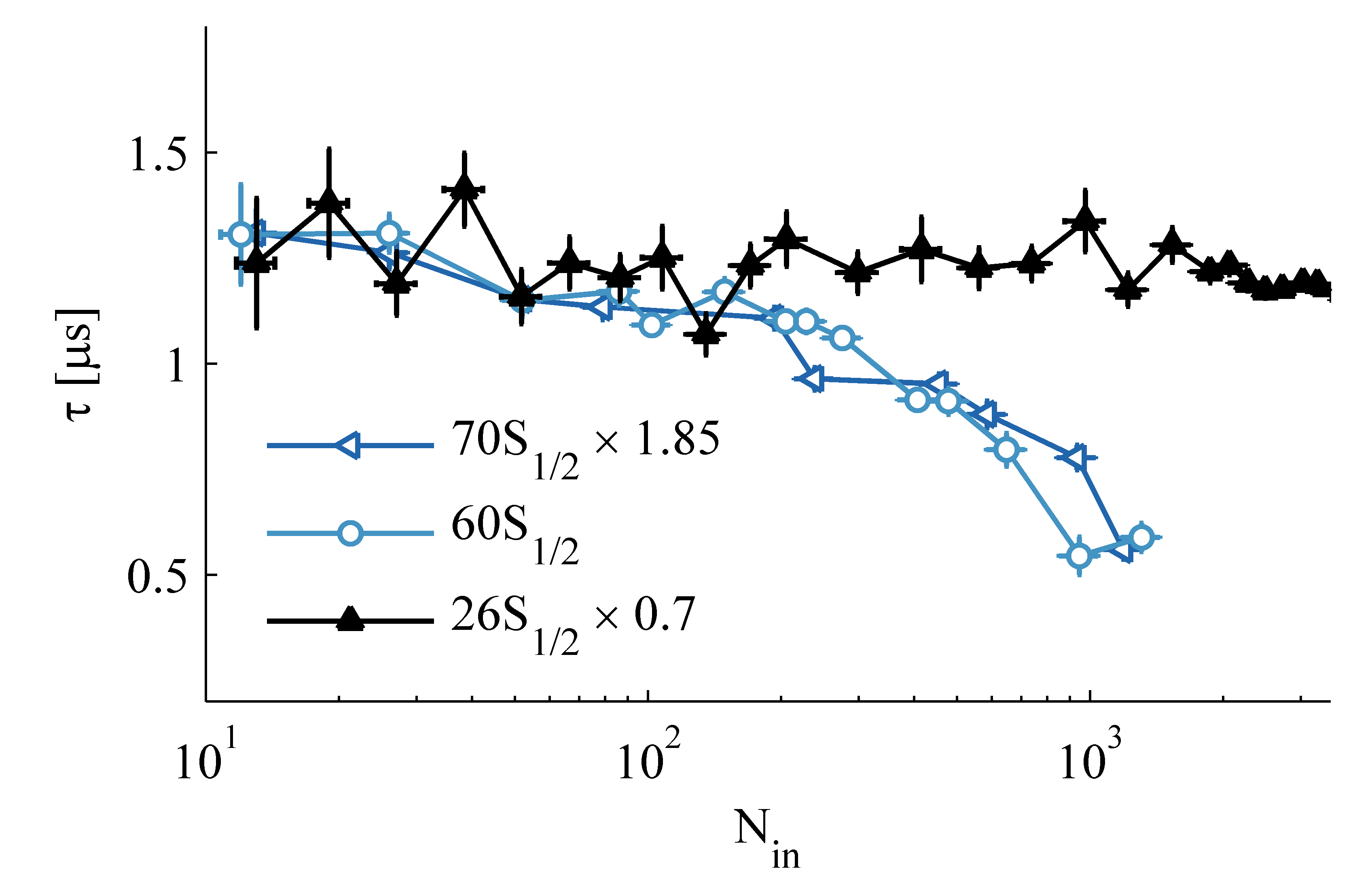}
	\caption{Coherence time as a function of input photon number. Comparison between the low-lying $26S_{1/2}$ state (black filled triangles) and the data for the interacting states $60S_{1/2}$ and $70S_{1/2}$ (empty circles and triangles respectively) shown in Fig. 4 of the main text. To compare the data, the coherence time for the $26S_{1/2}$ is multiplied by a factor 0.7 while for the $26S_{1/2}$ by a factor 1.85} 
	\label{fig:tau26S}
\end{figure}

Since the dipole-dipole interaction are negligible for low-lying states, we do not expect to see any input number of photon dependence of the coherence time, $\tau$. This is indeed the case, as we show in Fig. S\ref{fig:tau26S}. Following up the discussion in the last paragraph of the main text, for low lying Rydberg states the density of photons is lower than the density of super-atoms for the parameters range considered in the experiment. In this case the average distance between the stored Rydberg excitations, $r'$, would be $r' \gg r_{b}$, therefore $V(r')\ll \hbar \delta_{\rm EIT}$ and no Rydberg induced dephasing is expected. 

\section{Linear efficiency, T}
For deterministic photonic QIP, one requires an efficient and coherent mapping between light and a material which enables strong and controllable interactions among photons. In the main text of the paper we have rescaled the efficiencies by the linear efficiency, T, for the sake of highlighting the effects of the nonlinearity; here we present a longer discussion of the linear efficiency in the storage process.

In our experiment, the aforementioned mapping between light and our cold atomic ensemble -- EIT storage -- is characterized by the linear efficiency, $\displaystyle T = \lim_{N_{\rm in}\rightarrow 0} N_{\rm out}/N_{\rm in}$, which is limited by the optical depth of our cloud and by other sources of dephasing, such as the motion of the atoms, laser linewidth and stray fields. In particular, the dephasing limits the maximum achievable transparency on resonance and broadens the EIT peak. Together with finite OD, this diminishes the capability of fully compress the pulse inside the atomic cloud when performing storage (see \cite{Gorshkov07}), limiting $\displaystyle T$ at short storage time. In turn, the decay of the linear efficiency over time ($\tau$ at low $N_{\rm in}$ in Fig. 4 of the main text) is set by the dephasing only, which is dominated by stray fields in our experiment. 

\begin{figure}[ht]
	\includegraphics[width=\linewidth]{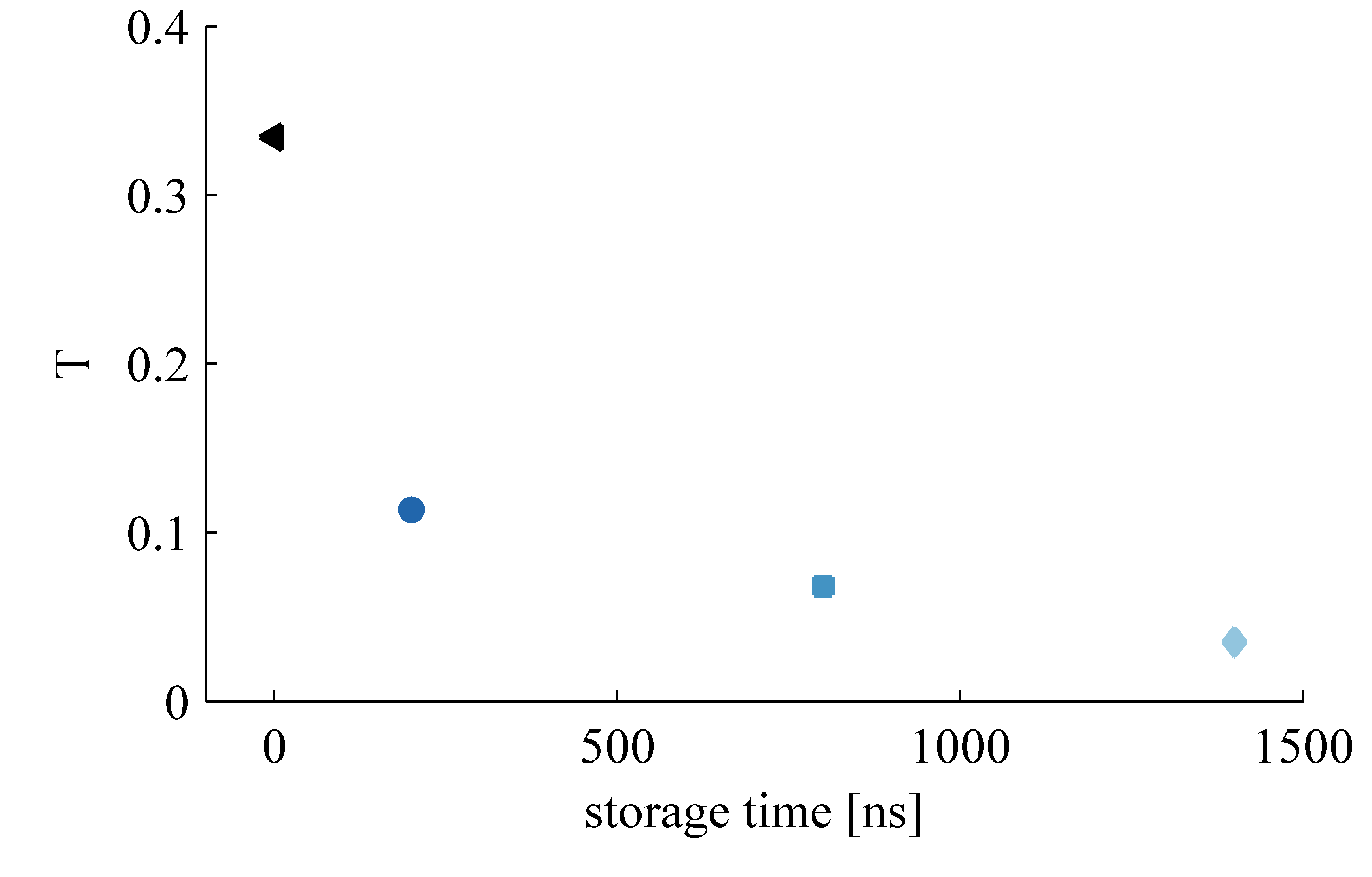}
	\caption{The linear efficiency T for the $\ket{70S_{1/2}}$ state is shown as a function of the storage time. Data are extracted from the fit of the curves in Fig. 2(a) in the main text. Error-bars are included in the markers.}  
	\label{fig:TvsStTime}
\end{figure} 

In Fig. S\ref{fig:TvsStTime} we show the linear efficiency T for the $\ket{70S_{1/2}}$ state for the cases shown in Fig. 2 of the main text. In our experiment, $\displaystyle T$ at short storage time is mostly limited by the OD. At higher OD, efficiencies near unity are possible albeit challenging \cite{Chen13,Hsiao16}. At longer storage time the decay of the efficiency is fundamentally limited by half of the lifetime of the Rydberg state, $\Gamma_r/2$, which increases with increasing principal quantum number as $\tau\propto n^{3}$. This limit might be approached in future experiment by compensating for stray fields, and by limiting the motion of the atoms using standard cooling and trapping techniques, such as an optical lattice.

\begin{figure}[h]
	\includegraphics[width=\linewidth]{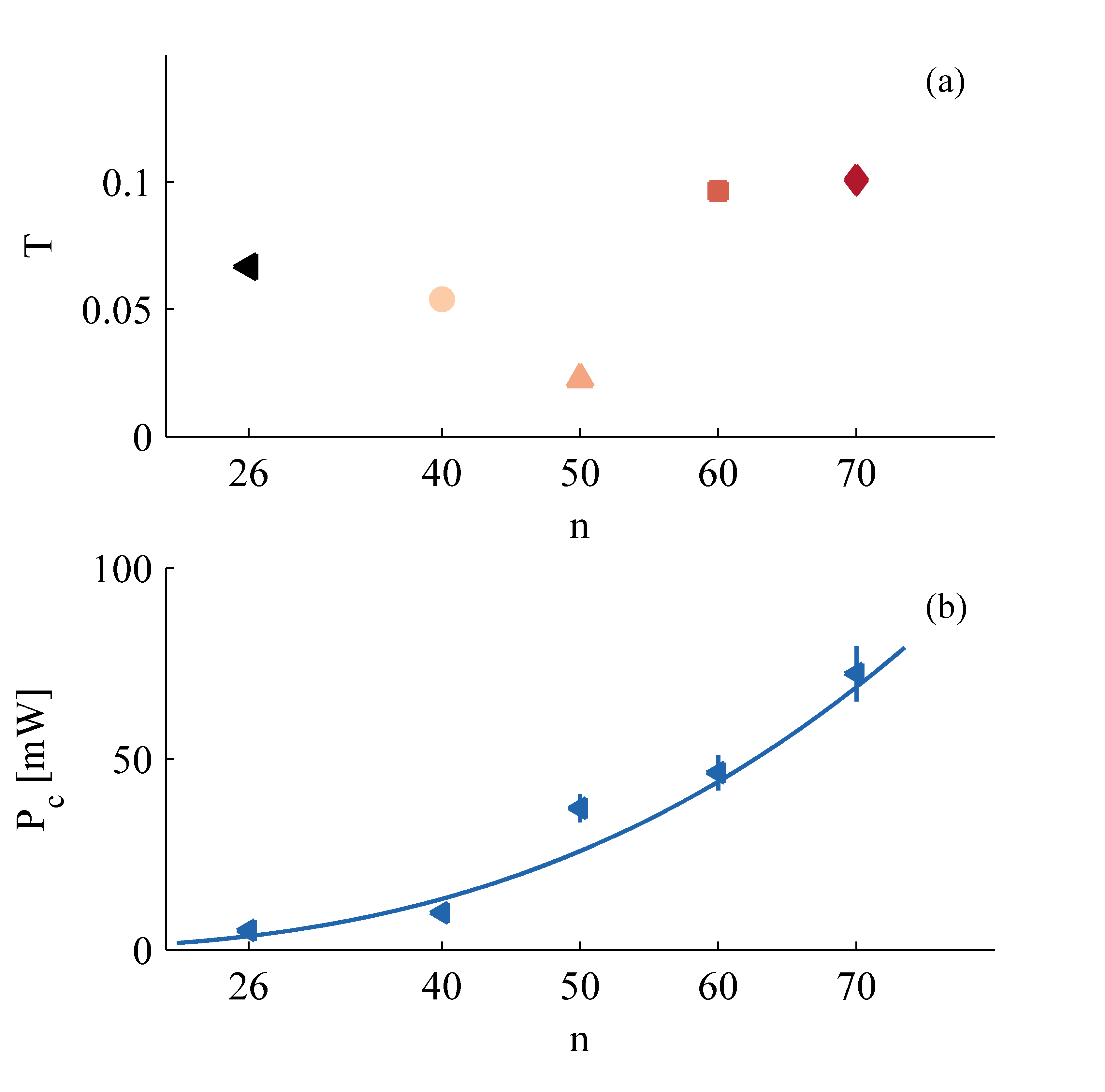}
	\caption{(a) Linear efficiency T for the curves in Fig. 2(b) of the main text. (b) Power of the coupling laser as a function of the principal quantum number $n$ used to keep $\Omega_c$ constant. The solid line is a fit with the function $P_{c}  = \alpha (n^*)^\beta$.}  
	\label{fig:Tvsn}
\end{figure}

Another limitation to T, either in the slow-light case and in the storage case, is the finite EIT transmission at the resonance frequency, $T_{0}$ in Eq. \eqref{eit:eq}. In order to compare different Rydberg states with similar condition, we maintained the coupling Rabi frequency $\Omega_{c}$ constant for the whole data set, resulting in a roughly constant T as a function of the principal quantum number $n$. This is shown in Fig. S\ref{fig:Tvsn}(a) where T as a function of $n$ is plotted for the same set of data as in Fig. 2(b). 

Maintaining $\Omega_{c}$ constant is achieved at cost of increasing the coupling laser power $P_{c}$. $\Omega_{c}$ is proportional to the dipole matrix element $d_{\rm n}$ of the $\ket{5P_{3/2}}\rightarrow\ket{nS_{1/2}}$ transition which scales as $d_{\rm n} \sim (n^*)^{-3/2}$, where $n^*$ is the principal quantum number corrected by the quantum defect theory. As results, the power needed to keep $\Omega_{c}$ constant scales as $P_{c} \sim (n^*)^3$. We explicitly show this in Fig. S\ref{fig:Tvsn}(b), where the data of the power used for the coupling laser are plotted as a function of $n$. A fit with the function $P_{c}  = \alpha (n^*)^\beta$ gives $\beta = 2.74 \pm 0.14$, with $n^* = n - \delta_{0}-\delta_{2}/\left(n-\delta_{0}\right)^2$, where $\delta_{0} = 3.1312$ and $\delta_{2}= 0.1787$.

In the paper we show that it is possible to enhance the non-linearity by performing storage. This only requires control of the dephasing sources and a high OD for an efficient light-matter coupling. In contrast, typical protocols using Rydberg polaritons require high OD per blockade (OD$_{\rm b}$) to implement strong photon-photon interactions. High OD$_{\rm b}$ can be achieved at high density of the atomic cloud and at high principal quantum number. However at high OD$_{\rm b}$ the interaction between the external Rydberg electron and the surrounding atoms acts as an additional source of decoherence \cite{Firstenberg2016}. Moreover, as shown in Fig. S\ref{fig:Tvsn}(b), reaching high $n$ might be experimentally challenging. From one side, the finite power of the coupling laser might be a limitation and from the other side -- as shown also in the next section --  higher energetic Rydberg states are more affected to coupling with external stray fields, which would limit the fidelity of the photon-photon interaction.    

\section{Coherence time}
In this section we overview the model that we employed to extract the coherence times $\tau$ in Fig. 4 of the main text. 

\begin{figure}[h]
	\includegraphics[width=\linewidth]{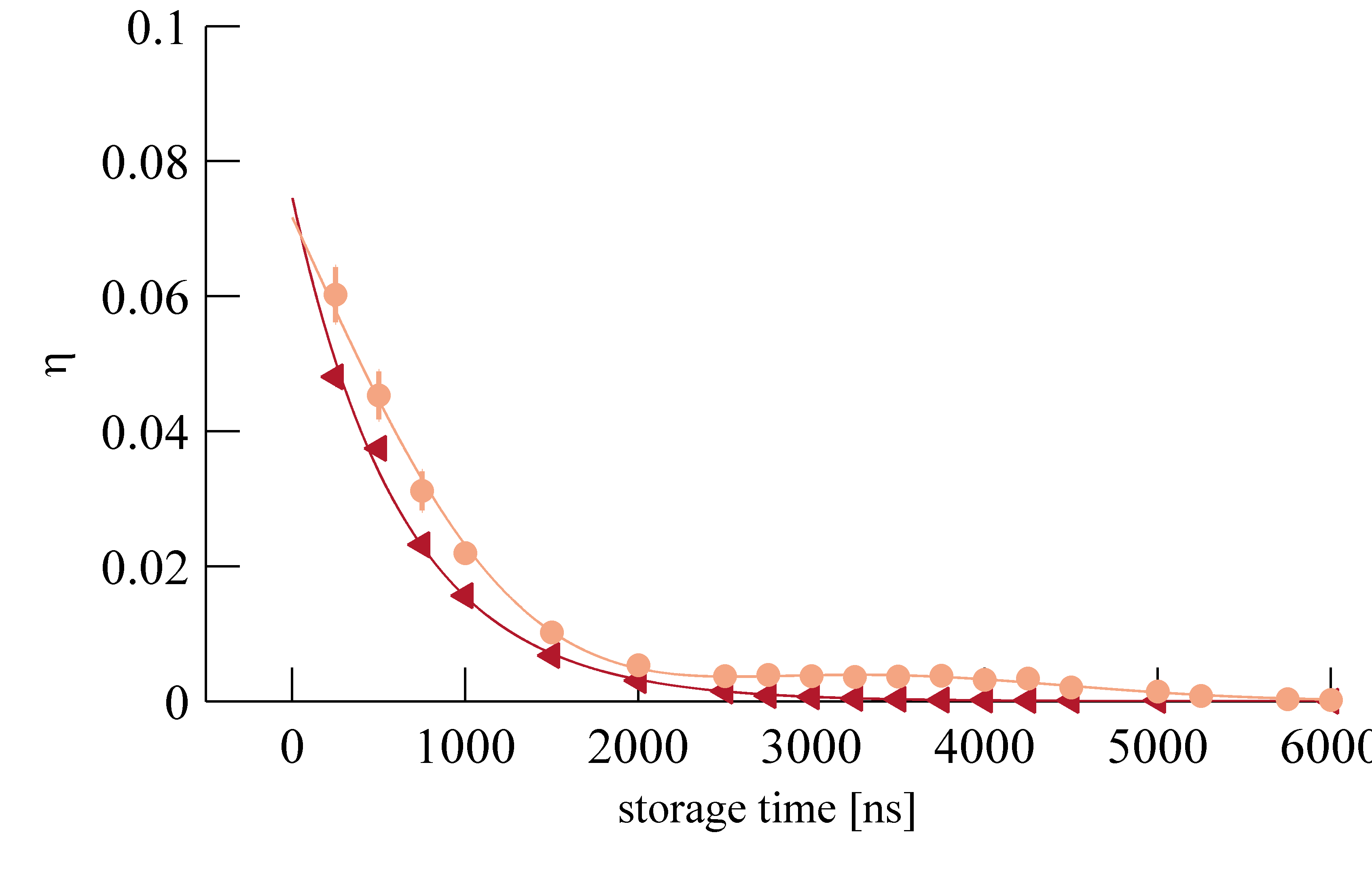}
	\caption{Storage efficiency $\eta$ as a function of the storage time for $N_{\rm in} = 26.4$ and $N_{\rm in} = 51.8$ for the states $60S_{1/2}$ (orange) and $70S_{1/2}$ (red) respectively. Solid line represent a fit with the model shown in the main text.}  
	\label{fig:dephasing}
\end{figure}

We measured the storage efficiency $\eta$ as a function of the storage time $t_{\rm s}$ for different $N_{\rm in}$ and for two Rydberg states $\ket{60S_{1/2}}$ and $\ket{70S_{1/2}}$. For low $N_{\rm in}$ the Rydberg dipole-dipole interaction is negligible. In this case, the decay of $\eta$ is due to other sources of dephasing, as atomic motion and stray fields. We observe an exponential decay of $\eta$ for both Rydberg states (see Fig. S\ref{fig:dephasing}) which suggest that atomic motion due to finite temperature is not our main source of dephasing. 

The storage efficiency for the state $\ket{60S_{1/2}}$ presents oscillations at a frequency $\Delta F = 231 \pm 1$ kHz. We attribute this oscillations to the hyperfine splitting of the Rydberg states. Due to our finite laser linewidth, we excite both hyperfine states $\ket{60S_{1/2}, F = 1}$ and $\ket{60S_{1/2}, F=2}$ which are separated by $\Delta F_{\rm theo} = 182.1$ kHz. To include this in our model we fit $\eta$ with the function
\begin{equation}
\eta = \eta_{0}e^{-t_{\rm s}/\tau} \left| p_{\rm F=1} +(1-p_{\rm F=1})e^{-2\pi \Delta F t_{\rm s}}\right|^2
\end{equation}   
where $p_{\rm F=1}$ is the probability to excite the $\ket{60S_{1/2}, F = 1}$ state. From the fit we can extract $\tau$ as well as $\eta_{0}$ and $\Delta F$. We attribute the difference between $\Delta F$ and $\Delta F_{theo}$ to undesired external electric field. 

The coherence time that we observe for the $\ket{70S_{1/2}}$ states at low $N_{\rm in}$ is 1.85 times shorter than for the $\ket{60S_{1/2}}$ state. This might be attributed to external electric field which affects more the coherence time of higher energetic states, since the polarizability scales as $\alpha \sim (n^*)^7$. Due to shorter coherence time, oscillations can not be observe for the $\ket{70S_{1/2}}$ state in our experiment, therefore the storage efficiency display a simple exponential decay as $\eta = \eta_{0}\rm exp(-t_{\rm s}/\tau)$.

\end{document}